# Small bandgap features achieved in atomically precise 17-atom-wide armchair-edged graphene nanoribbons

**Authors**

Junichi Yamaguchi[1,*], Hironobu Hayashi[2,*], Hideyuki Jippo[1], Akitoshi Shiotari[3], Manabu Ohtomo[1], Mitsuhiro Sakakura[2], Nao Hieda[2], Naoki Aratani[2], Mari Ohfuchi[1], Yoshiaki Sugimoto[3], Hiroko Yamada[2] & Shintaro Sato[1]

[1]Fujitsu Laboratories Ltd. and Fujitsu Limited, 10-1 Morinosato-Wakamiya, Atsugi, Kanagawa 243-0197, Japan

[2]Division of Materials Science, Nara Institute of Science and Technology, 8916-5 Takayama-cho, Ikoma, Nara 630-0192, Japan

[3]Department of Advanced Materials Science, The University of Tokyo, Kashiwa, Chiba 277-8561, Japan

[*]These authors contributed equally to this work.

Correspondence and requests for materials should be addressed to J.Y. (email: yamaguchi.j@fujitsu.com) and/or H.Y. (email: hyamada@ms.naist.ac.jp)

**Abstract**

Graphene nanoribbons (GNRs) synthesized using a bottom-up technique potentially enable future electronic devices owing to the tunable electronic structures depending on the well-defined width and edge geometry. For instance, armchair-edged GNRs (AGNRs) exhibit width-dependent bandgaps. However, the bandgaps of AGNRs synthesized experimentally thus far are relatively large, well above 1 eV. Such a large bandgap may deteriorate device performances due to large Schottky barriers and carrier effective masses. We describe the bottom-up synthesis of AGNRs with a smaller bandgap using dibromobenzene-based precursors. Two types of AGNRs with different widths of 17 and 13 carbon atoms were synthesized on Au(111), and their atomic and electronic structures were investigated by scanning probe microscopy and spectroscopy. We reveal that the 17-AGNRs has the smallest bandgap as well as the smallest electron/hole effective mass among bottom-up AGNRs reported thus far. The successful synthesis of 17-AGNRs is a significant step toward the development of GNR-based electronic devices.

Quasi-one-dimensional strips of graphene with nanoscaled width, so-called graphene nanoribbons (GNRs), exhibit unique electronic and magnetic properties that do not appear in two-dimensional graphene[1-4]. These properties can emerge from the structural boundary conditions imposed by the atomically precise width and edge structure in GNRs. Two representative types of GNRs are armchair-edged GNRs (AGNRs) and zigzag-edged GNRs (ZGNRs). AGNRs show sizable gapped electronic states established by the quantum confinement and edge effects[5-8]. In graphene-based electronics, AGNRs with finite bandgaps at room temperature have attracted much attention as a reliable semiconducting channel for field-effect transistors (FETs). Numerous top-down approaches have been conducted to fabricate GNRs[9-11]. Nevertheless, such GNRs typically had uncontrollable edge geometries with defects, exhibiting poor transport properties as a transistor channel.

For fabricating atomically precise GNRs, an advanced bottom-up synthesis technique has recently been proposed and demonstrated, which uses on-surface-assisted polymerization and subsequent cyclodehydrogenation of precursor monomers on metal substrates[12]. This technique has yielded atomically precise AGNRs[12-14] and ZGNRs[15]. In particular, since the bandgap of AGNRs can be tuned by changing the width, the bottom-up synthesis for $N$-AGNRs (where $N$ is the width in the number of rows of carbon atoms across the AGNRs) has been widely conducted. Following the pioneering 7-AGNRs on Au(111)[12], various $N$-AGNRs have been reported, such as 3-AGNRs[16, 17], 5-AGNRs[18], 9-AGNRs[19] and 13-AGNRs[20], as well as the derivative AGNRs incidentally cyclodehydrogenated along the width direction (e.g., 14-, 18- and 21-AGNRs)[21, 22].

As theoretically predicted, $N$-AGNRs can be categorized into three subfamilies with $N$ = $3p$, $3p+1$ and $3p+2$ ($p$ is a natural number), in which each electronic structure varies depending on the structural boundary conditions[6, 7]. For evaluating the quasiparticle bandgaps of AGNRs accurately, the first-principles calculations by considering many-body perturbation theory ($GW$ approximation[23], where $G$ is the Green's function and $W$ is the screened Coulomb interaction) have been carried out beyond the framework of the density functional theory (DFT)[7]. According to the calculations, the gap size $\Delta$ decreases with increasing $N$ (i.e., the width) within each subfamily and follows the relation among the subfamilies with the same $p$, namely $\Delta^{3p+1} > \Delta^{3p} > \Delta^{3p+2}$. For instance, with the $GW$ calculations for freestanding $N$-AGNRs[7, 24], the quasiparticle gap $\Delta_{GW}$ is predicted to be 3.80 and 2.25 eV for $N$ = 7 and 13 ($3p+1$), respectively, 2.16 eV for $N$ = 9 ($3p$) and 1.75 eV for $N$ = 5 ($3p+2$). Previous studies on FETs using bottom-up AGNRs with widths of $N$ = 7, 9 and 13 presented their electrical transport characteristics[25, 26]. However, in such GNR-FETs, the intrinsic transport properties of AGNRs were obscured by large Schottky barrier resistance between AGNRs and metal contacts. This is because the bandgaps in the above AGNRs are large, substantially exceeding 1 eV. The transport characteristics can be improved through the use of wider AGNRs since they are expected to have lower Schottky barriers and smaller effective masses arising from their small bandgap features. In fact, in carbon nanotubes (CNTs) with bandgaps smaller than 1 eV, much better transport characteristics have been reported[27, 28]. Hence, AGNRs belonging to the $3p+2$ subfamily, which exhibit the smallest bandgaps among the three subfamilies, have considerable potential to be exploited in GNR-FETs. GNR-FETs can have an advantage over CNT-FETs because the structure and, therefore, properties of GNRs can be precisely controlled using the bottom-up synthesis, in contrast to the case of CNTs.

We discuss the bottom-up synthesis of the well-structured 17-AGNRs ($3p+2$) as well as 13-AGNRs ($3p+1$) on Au(111) in ultra-high vacuum by using two types of dibromobenzene-based precursor monomers. To systematically control the width of the AGNRs, in the monomers, two anthracene units and two naphthalene units are introduced into the dibromobenzene for the 17- and 13-AGNRs, respectively. Although 13-AGNRs synthesized with a different precursor monomer were already reported[20], we show the consistency between our 13-AGNRs and the previous ones. Characterizations of the atomic and electronic structures of the 17- and 13-AGNRs on Au(111) are carried out by combining in-situ scanning tunneling microscopy and spectroscopy (STM/STS) and ex-situ non-contact atomic force microscopy (nc-AFM). We find that the experimentally obtained electronic structures in both AGNRs are consistent with the accordingly corrected quasiparticle states predicted from the *GW* calculations. We reveal that the 17-AGNRs have a bandgap of 0.19 eV on Au(111), which is consistent with a theoretically obtained bandgap of 0.63 eV for a freestanding 17-AGNR[24]. As far as we know, this is the first demonstration of the synthesis of GNRs having a bandgap smaller than 1 eV in a controlled manner.

## Results

**Synthesis of 17- and 13-AGNRs.** To control the width of bottom-up AGNRs, we developed two dibromobenzene-based precursor monomers shown in Fig. 1. The monomers 1,2-bis-(2-anthracenyl)-3,6-dibromobenzene (BADBB) for the 17-AGNRs (Fig. 1a) and 1,2-bis-(2-naphthalenyl)-3,6-dibromobenzene (BNDBB) for the 13-AGNRs (Fig. 1b) were successfully obtained through multi-step organic synthesis (see Supplementary Note 1). Annealing a Au(111) surface up to 250 ℃ induced the detachment of bromine atoms from BADBB and BNDBB and extended one-dimensional 17- and 13-polymers were formed by aryl−aryl coupling. The covalent bonds between the monomers can form only when they are rotated 180 degrees to each other, as in the synthesis of 9-AGNRs[19, 29]. Upon further annealing to 400 °C, thermally induced cyclodehydrogenation in the polymers led to the formation of the 17- and 13-AGNRs. The BADBB or BNDBB monomers were individually deposited onto a clean Au(111) surface maintained at room temperature by thermal sublimation under ultra-high vacuum conditions. After room-temperature deposition, it is confirmed using STM that both BADBB and BNDBB monomers tend to form self-assembled islands on Au(111) (Supplementary Fig. 11a, b). Their apparent heights determined from STM are 0.24 nm for BADBB and 0.22 nm for BNDBB. To synthesize long and low-defective AGNRs, the samples were step-wisely annealed from room temperature to 400 °C in increments of 50 °C, instead of a conventional two-step annealing at 200 and 400 °C for previous bottom-up GNRs[12, 19, 20].

We successfully observed the formations of the polymers and AGNRs using STM. Figure 2a shows a large-scale STM topographic image of 17-polymers on Au(111) after annealing at 250 °C. The 17-polymers also assemble into the extended islands as with the BADBB monomers after room-temperature deposition. The apparent height of the 17-polymers, however, slightly increases to 0.34 nm compared with 0.24 nm of the BADBB monomers. This increase is attributed to the sterically induced out-of-plane conformation of the anthracene side units due to the polymerization. A small-scale STM image of the 17-polymers is presented in Fig. 2b, together with a structural model obtained from DFT calculations for a freestanding 17-polymer. As seen from the STM image, the protrusions, derived from the sterically hindered anthracene side units, align with a periodicity of 0.86±0.05 nm along the polymer axis. From the calculations, the predicted periodicity of a repeat unit is 0.90 nm, which is in good agreement with the experimental value. In addition, the simulated STM image of the 17-polymer (Fig. 2c) can well reproduce the experimentally observed periodic protrusions.

In the case of the 13-polymers, the large-scale STM image (Fig. 2d) indicates that they exhibit a similar island formation to that of the 17-polymers. The apparent height increases from 0.22 nm for the BNDBB monomers to 0.32 nm for the 13-polymers. The small-scale STM image and DFT simulated image for the 13-polymers are presented in Fig. 2e and f. The periodicity of the protrusions, arising from the sterically hindered naphthalene side units, is estimated to be 0.95±0.08 nm along the polymer axis. This value agrees with the periodicity of 0.90 nm predicted from the calculations, and the experimental STM image closely matches the simulated image.

Further annealing to 400 °C results in a complete planarization of the protrusive polymers and leads to fully conjugated AGNRs, which was observed as a reduced apparent height of 0.18 nm in both 17- and 13-AGNRs (Fig. 3a, d). Each width is estimated to be 2.5 and 1.9 nm with an accuracy of ±0.1 nm in the 17- and 13-AGNRs, respectively. The dimensional feature of these 13-AGNRs is consistent with that of previous ones synthesized with a different precursor monomer[20]. The high-resolution STM images of the 17- and 13-AGNRs are shown in Fig. 3b and e, together with their structural models of freestanding 17- and 13-AGNRs calculated by DFT. The observed molecular structures exhibit the edge periodicity of 0.42±0.03 nm along the ribbon axis in both AGNRs. The experimental periodicity is in good agreement with the simulated value (0.43 nm). The observed ring-like shapes within the AGNRs also agree with the simulated STM images (Fig. 3c for a 17-AGNR and f for a 13-AGNR).

To check further structural details of the 17- and 13-AGNRs, we carried out ex-situ nc-AFM imaging with CO-functionalized tips. Samples were transferred through the air into the AFM measuring chamber and subsequently annealed at about 400 °C under ultra-high vacuum conditions to remove contaminations on the surface adsorbed during air exposure. Figure 4a shows the frequency shift image of the 17-AGNR. For more visibility, the corresponding Laplace filtered image is depicted in Fig. 4b. The Laplace filtered image directly reveals the width consisting of 17 carbon atoms corresponding to the expected 17-AGNR structure. For the 13-AGNR, the frequency shift image and Laplace filtered one are presented in Fig. 4c and d, respectively. The width consisting of 13 carbon atoms is also confirmed in Fig. 4d. Note that despite ex-situ sample preparations, we were able to obtain the bond-resolved AFM images of bottom-up GNRs for the first time. Our results indicate that nc-AFM has versatile usability for probing atomic structures, even when preparing samples under ex-situ conditions. In addition to the STM and nc-AFM characterizations, the quality of the 17- and 13-AGNRs on a larger scale were investigated by X-ray photoelectron and Raman spectroscopies. The detailed results are described in Supplementary Notes 2 and 3.

**Electronic structures of 17- and 13-AGNRs.** To acquire information about the electronic structures of the 17- and 13-AGNRs, we carried out STS measurements. In the Tersoff-Hamann approximation[30], a differential conductance ($dI/dV$) obtained by STS is proportional to the local density of states (LDOSs) at the position of the STM tip. The LDOSs in the 17- and 13-AGNRs were characterized using the $dI/dV$ point spectra, as shown in Fig. 5a, c. All spectra were obtained after calibrating the STM tip by confirming the appearance of the so-called Shockley surface states on Au(111) around $V_s \sim -0.4$ V (Ref. 31).

By comparing the d$I$/d$V$ spectrum obtained on the edge of the 17-AGNR (red line) to that obtained on the bare Au(111) surface (black dotted line) in Fig. 5a, we notice the ribbon-related peaks centered at −0.09±0.02 V (occupied states) and 0.10±0.02 V (unoccupied states), which can be regarded as the valence band maximum (VBM) and conduction band minimum (CBM), respectively, since they bracket the Fermi level $E_F$ (at $V_s = 0$ V). From the energy difference between these peaks, the experimental energy gap is estimated to be $\Delta_{STS} = 0.19±0.03$ eV for the 17-AGNR on Au(111). In the $GW$ calculations, the quasiparticle gap of a freestanding 17-AGNR is predicted to be $\Delta_{GW} = 0.63$ eV (Ref. 24). In general, the values of $\Delta_{STS}$ determined by STS in GNRs on metal surfaces are significantly underestimated to those of $\Delta_{GW}$ for freestanding GNRs. The reduction in $\Delta_{STS}$ to $\Delta_{GW}$ is caused by the substrate-induced weakening of the electrostatic potential due to the long-range screening effects[22]. Therefore, the $\Delta_{GW}$ values should be corrected by considering the substrate screening when compared with those of $\Delta_{STS}$. According to Ref. 32, the quasiparticle gaps of substrate-supported GNRs have been corrected using an advanced image-charge model that includes the substrate screening as well as the internal screening of the GNRs. The renormalized quasiparticle gap of 17-AGNRs is predicted to be $\Delta_{GW'} = 0.20$ eV (Ref. 22), which is in good agreement with our experimental energy gap. In fact, the renormalized gap was computed with a model of 17-AGNRs adsorbed on a $Au_3Si$ monolayer intercalated between the ribbon and Au(111) substrate[22]. Intercalating a semiconducting or insulating layer into the GNR−substrate interface increases the relative distance between the GNR and image-plane of the substrate. The increase in the distance leads to the reduction in the image-charge corrections, expanding the renormalized gap. However, in wide GNRs, such as 17-AGNRs, the internal screening of GNRs is, owing to their large polarizabilities, more dominant than the external screening from the substrate[32]. In such cases, the change in the renormalized gap depending on the presence or absence of the intercalated layer is almost negligible. Therefore, we can safely conclude that the agreement of our experimental energy gap with the theoretical renormalized gap is consistent.

The spatial distributions of the electronic structure of the 17-AGNR were experimentally explored by d$I$/d$V$ mapping at different sample biases (Fig. 5b). The d$I$/d$V$ maps obtained at $V_s = 2.0$ and −1.0 V exhibit significant LDOSs along the two edges of the ribbon. On the other hand, in the d$I$/d$V$ maps at $V_s = 0.1$ and −0.09 V corresponding to the energies of CBM and VBM, the enhancement of the d$I$/d$V$ intensity along the edges is partially suppressed because of the oscillatory contrast derived from the quantum interference of the Au(111) surface states at these biases[33].

Figure 5c illustrates the d$I$/d$V$ spectra taken on the edge of a single 13-AGNR and Au(111) surface. The spectrum of the 13-AGNR shows two prominent peaks centered at −0.06±0.02 and 1.28±0.02 V. These peak positions are similar to the energy positions of the VBM and CBM previously measured by STS for 13-AGNRs on Au(111)[20]. The energy gap is estimated to be $\Delta_{STS}$ = 1.34±0.03 eV from the energy difference between those peaks. This energy gap is also consistent with the previous one (1.4±0.1 eV)[20]. With the $GW$ calculations, the quasiparticle gap is predicted to be $\Delta_{GW}$ = 2.25 eV for a freestanding 13-AGNR[24], and the renormalized gap of the 13-AGNR supported by the Au(111) substrate is corrected as $\Delta_{GW}$' = 1.29 eV using the advanced image-charge model[32]. The theoretical gap shows good agreement with our experimental one.

For this 13-AGNR, the d$I$/d$V$ maps measured at the characteristic sample biases are depicted in Fig. 5d. The d$I$/d$V$ maps clearly show significant LDOSs along the edges at the CBM and VBM, as well as in the occupied state (at $V_s$ = −1.0 V). In contrast, there is no prominent d$I$/d$V$ intensity on the ribbon in the map at $V_s$ = 0.7 V since this sample bias corresponds to the energy within the bandgap. These LDOS behaviors in the 13-AGNR by varying the sample bias were also observed in previous STS measurements[20].

**Comparison between experimental and theoretical band structures.** For more detailed discussions on the dispersion of the electronic states of the 17- and 13-AGNRs, we carried out Fourier-transformed STS (FT-STS) measurements. Figure 6a shows a series of d$I$/d$V$ spectra measured along one armchair edge of a single 17-AGNR at intervals $\delta x$ = 0.11 nm (see Supplementary Fig. 14a for the measured 17-AGNR). In the LDOS [i.e., d$I$/d$V$ ($V$, $x$)] map, while the standing wave patterns derived from scattering at the termini of the ribbon are clear in the unoccupied states ($V_s$ > 1.3 V) and occupied states ($V_s$ < −0.5 V), the weak patterns are also clear in the vicinity of $E_F$. Moreover, the absence of the LDOSs within the bandgap can be confirmed throughout the measuring positions. The LDOS maps of a single 13-AGNR in the unoccupied and occupied states are presented in Fig. 6c and d (the measured 13-AGNR is shown in Supplementary Fig. 14b, c). Note that the LDOSs of the 13-AGNR are absent in the region from $V_s$ ~ 1 V to $E_F$, reflecting the bandgap.

To obtain the electronic band dispersion, we carried out a discrete Fourier transform of d$I$/d$V$ ($V$, $x$) in real space to reciprocal space (see detailed procedures in Ref. 34). Figure 6b presents the FT-LDOS map of Fig. 6a for the 17-AGNR in the range of the wave vector $0 \leq k$ (= $q/2$) $\leq \pi/a$ corresponding to the first Brillouin zone of the ribbon. The FT-LDOS maps of the 13-AGNR in the unoccupied and occupied states are shown in Fig. 6e and f. In both 17- and 13-AGNRs, the FT-LDOS maps reveal the appearance of dispersing bands and the bandgaps. It is, however, difficult to discuss the respective bands by resolving each other due to the insufficient signal-to-noise ratio. This experimental problem could be due to the finite tip size and a deviation from a constant tip−sample distance (approximately 0.1 nm)[19, 34]. We thus computed the quasiparticle bands of freestanding 17- and 13-AGNRs using the $GW$ calculations to compare them with the experimental band dispersions. Figure 6g and h show the calculated band structures of the 17- and 13-AGNRs. From our calculations, the quasiparticle gap of these AGNRs is predicted to be $\Delta_{GW} = 0.72$ and 2.01 eV, which is in acceptable agreement with the previous calculation results[24].

   As mentioned above, the $GW$ calculations overestimate the bandgap of GNRs compared with the STS experiments. We thus attempted to rigidly shift the $GW$ quasiparticle bands to compare them with the experimental band dispersions (see red dotted curves in Fig. 6b, e, f). All quasiparticle CBs and VBs are shifted to match the onset of the quasiparticle CB and VB with the energies of the experimental CBM and VBM (Fig. 5a, c). Surprisingly, the rigid-shifted quasiparticle bands reasonably reproduce the experimentally observed band structures in both 17- and 13-AGNRs, although each quasiparticle band is not corrected by considering the image-charge effects.

   It is difficult to accurately estimate the effective masses by using these experimental data. Thus, as an indirect comparison, we carried out a parabolic least-squares fit [$E(k) = E(0) + \hbar^2 k^2/2m^*$] to the $GW$ quasiparticle CB and VB of these 17- and 13-AGNRs (depicted with blue dotted curves in Fig. 6g, h). The respective effective masses are estimated as $m^*_{CB} = m^*_{VB} = 0.06\ m_e$ in the 17-AGNR and $m^*_{CB} = 0.14\ m_e$ and $m^*_{VB} = 0.13\ m_e$ in the 13-AGNR (where $m_e$ is the free electron mass). Given the reasonably good agreement between the band structures obtained experimentally and theoretically, the actual effective mass is expected to be close to the $GW$ values. Notably, the small effective masses of both electrons and holes ($m^* \sim 0.06\ m_e$), as well as the small bandgap ($\Delta_{STS} = 0.19$ eV), were obtained in the 17-AGNR on Au(111). These values are substantially smaller than $m^* \sim 0.4\ m_e$ and $\Delta_{STS} = 2.4$ eV for 7-AGNRs[34] and $m^* \sim 0.1\ m_e$ and $\Delta_{STS} = 1.4$ eV for 9-AGNRs[19], which have been previously exploited in prototypes of GNR-FETs. The electron/hole effective mass in the 17-AGNR is even smaller than those in GaAs and InP (e.g., Ref. 35). Our results indicate that excellent devices including transistors can be obtained using 17-AGNRs in the near future.

## Conclusions

We discussed the synthesis of AGNRs with different widths of $N = 17$ and 13 on Au(111) by using the bottom-up technique with two types of dibromobenzene-based precursor monomers. STM and nc-AFM observations revealed their widths and armchair-edged structures with atomic precision. The local electronic structures characterized by STS allow for the determination of their bandgaps. The experimentally determined gaps of the 17- and 13-AGNRs on Au(111) are in fair agreement with the *GW* quasiparticle gaps corrected by considering the substrate screening. Furthermore, FT-STS provides the electronic band dispersions, which reasonably match the simply rigid-shifted *GW* quasiparticle bands in both types of AGNRs. We further found that the 17-AGNRs on Au(111) have a small bandgap of 0.19 eV and small effective masses of ~0.06 $m_e$ for electrons and holes. We expect that the 17-AGNRs successfully synthesized in this study will pave the way for the development of GNR-based electronic devices.

## Methods

**Sample preparation.** Experiments were conducted under ultra-high vacuum conditions (base pressure < $5 \times 10^{-8}$ Pa) with a low-temperature STM from ScientaOmicron. A Au(111) single crystal (MaTeck, Germany) and Au(111) epitaxial films on mica (Phasis, Switzerland) were used as substrates for the synthesis of the 17- and 13-AGNRs (we obtained similar results regardless of the types of substrates). Atomically clean Au(111) surfaces were prepared by repeated cycles of Ar ion sputtering and annealing at 500 °C. BADBB and BNDBB monomers (for details on the synthesis and characterization, see Supplementary Note 1) were thermally evaporated onto the clean Au(111) surface held at room temperature from a quartz crucible heated to 180 °C for BADBB and 120 °C for BNDBB, resulting in a deposition rate of ~1 Å/min in both monomers. The coverage of both monomers was controlled to be sub-monolayer, as determined from STM (Supplementary Fig. 11a, b). After deposition, the surface temperature was step-wisely ramped (<5 °C/min) from room temperature to 400 °C in five steps to 200, 250, 300, 350 and 400 °C. Each temperature step was held for 30 min at or less than 350 °C and for 2 h at 400 °C. Long and low-defective AGNRs were obtained owing to this step-wise annealing.

**Imaging and Spectroscopy.** STM measurements were conducted in the constant-current mode under $3 \times 10^{-9}$ Pa at a sample temperature of 77 K or 5 K. The temperature and scanning parameters are described in each figure caption. An electrochemically etched W tip was used for topographic and spectroscopic measurements. For STS measurements, all d*I*/d*V* signals were recorded at 5 K using a lock-in amplifier with a sinusoidal voltage of 10 mV (r.m.s) and frequency of 463 Hz. The d$I$/d$V$ spectra and maps were acquired under the open-feedback and constant-current conditions, respectively.

The nc-AFM experiments were conducted in another ultra-high vacuum chamber under $4\times10^{-9}$ Pa at a sample temperature of 5 K. Samples [17- and 13-AGNRs grown on Au(111)/mica substrates] mounted in the chamber were degassed at 200 °C for 4 h, then annealed at 400 °C for 20 min for the 17-AGNRs and at 420 °C for 30 min for the 13-AGNRs to remove impurities on the surface. For frequency modulation AFM, a tuning fork with an etched W tip was used as a force sensor (resonance frequency of 21.3 kHz, spring constant of 1800 N/m, quality factor of $1–5\times10^{4}$). To obtain high-resolution AFM images, a CO molecule coadsorbed on the surfaces at 6 K was picked up to attach to the tip apex[36]. Frequency shift was measured in the constant-height mode. For the AFM image of the 17-AGNRs (13-AGNRs), the tip height was 0.07 nm higher (0.03 nm lower) than the setpoint determined from STM at $V_s$ = 30 mV and $I_t$ = 20 pA over the bare Au surface. All scanning images were edited using SPIP software (Image Metrology, Denmark).

**Theoretical calculations.** Geometry optimizations for the 17- and 13-AGNRs were carried out using DFT within the generalized gradient approximation using the Perdew-Burke-Ernzerhof functional[37] for the exchange−correlation function as implemented in the OpenMX package (http://www.openmx-square.org/). The electron−ion interaction is described by norm-conserving pseudopotentials[38] under partial core correction[39]. Pseudo atomic orbitals (PAOs) centered on atomic sites are used as the basis function set[40]. The PAO basis functions are specified by C6.0-$s2p2d1$ and H5.0-$s2p1$. For example, C6.0-$s2p2d1$ indicates the PAO of the carbon atom with the cutoff radii of 6.0 Bohr and with two $s$, two $p$ and one $d$ component. The van der Waals corrections were included with a semiempirical DFT-D2 method[41]. Simulated STM images were obtained by the Tersoff-Hamann theory[30]. Partial charge density was calculated in an energy window $U$ measured from the chemical potential (shown in each figure caption) and visualized using WSxM software[42]. The lattice constant of the AGNRs was 4.30 Å, and the geometries were optimized under a three-dimensional periodic boundary condition with a criterion of 5.14 $\times 10^{-3}$ eV/Å for forces on atoms.

Quasiparticle band structure calculations were computed in the $GW$ approximation using the Berkeley$GW$ package[23, 43]. The electronic structure from DFT was recalculated using 60 Ry plane-wave cutoff and 64 $k$-points in the first Brillouin zone. The quasiparticle energies are determined by considering the lowest 29 and 45 unoccupied conduction bands for the 17- and 13-AGNRs, respectively. The static dielectric matrix $\varepsilon$ was calculated in the random phase approximation with 8 Ry cutoff for the plane-wave basis. The dynamical electronic screening is captured using the general plasmon pole model[23].

## Acknowledgments

This work was supported by CREST JST (no. JPMJCR15F1) and MEXT/JSPS KAKENHI grant nos. 26105004, 16H02286, 18K14190 and 18H01807. We thank Dr. Atsushi Ando of the National Advanced Industrial Science and Technology for supporting the Raman experiments. We also thank Prof. Riichiro Saito of Tohoku University for helpful advice on theoretical interpretations of the Raman spectra.

## Author contributions

J.Y. synthesized the AGNRs, conducted the STM/STS measurements and carried out the data analysis. A.S. and Y.S. conducted the nc-AFM imaging. H.H., M.S. and N.H. synthesized the precursor monomers. N.A. conducted the single-crystal X-ray analysis. J.Y., H.J. and M.O. conducted the theoretical simulations. S.S., H.Y. and M.O. supervised the project. J.Y., H.H., H.J. and A.S. wrote the manuscript, and all authors discussed the results and commented on the manuscript.

**Figures**

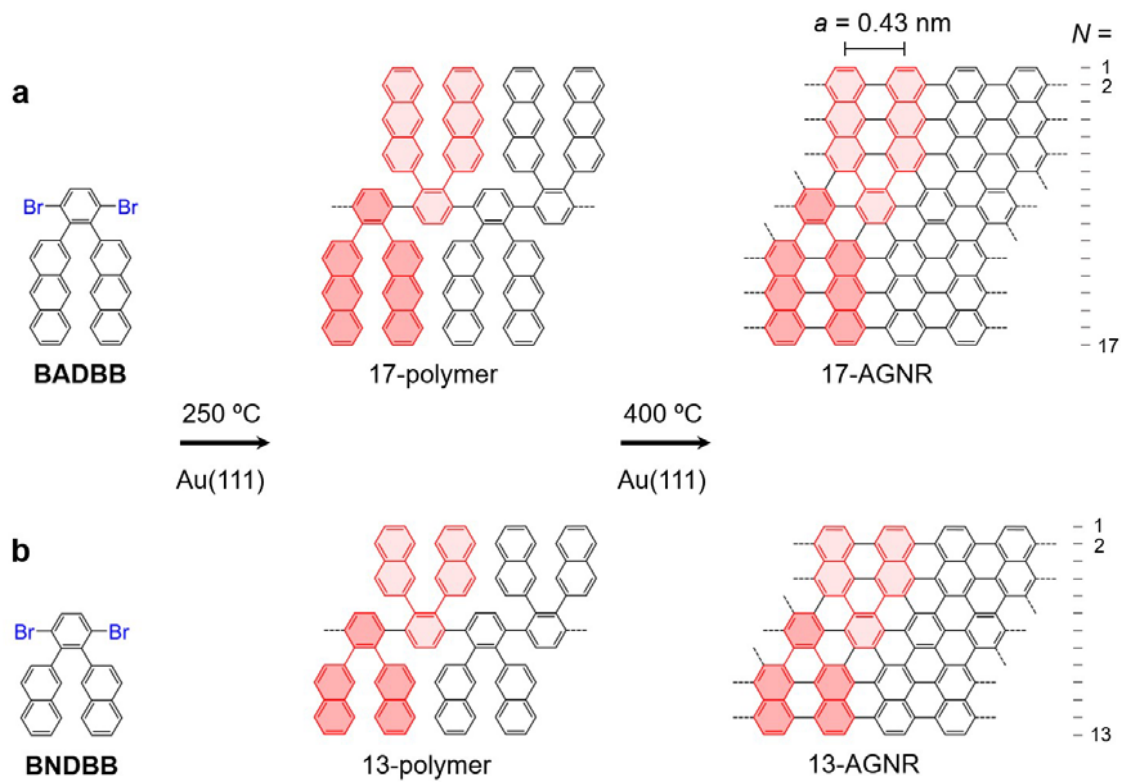

**Figure 1 | Bottom-up synthesis of 17- and 13-AGNRs.** (**a, b**) Schematic drawing of on-surface chemical reactions for (**a**) 17- and (**b**) 13-AGNRs on Au(111). 17- and 13-polymers are formed via dehalogenation and aryl−aryl coupling of BADBB and BNDBB monomers up to 250 °C, then cyclodehydrogenation of 17- and 13-polymers at 400 °C leads to formation of 17- and 13-AGNRs.The relevant lattice parameter of AGNRs is described as $a = 0.43$ nm in (**a**).

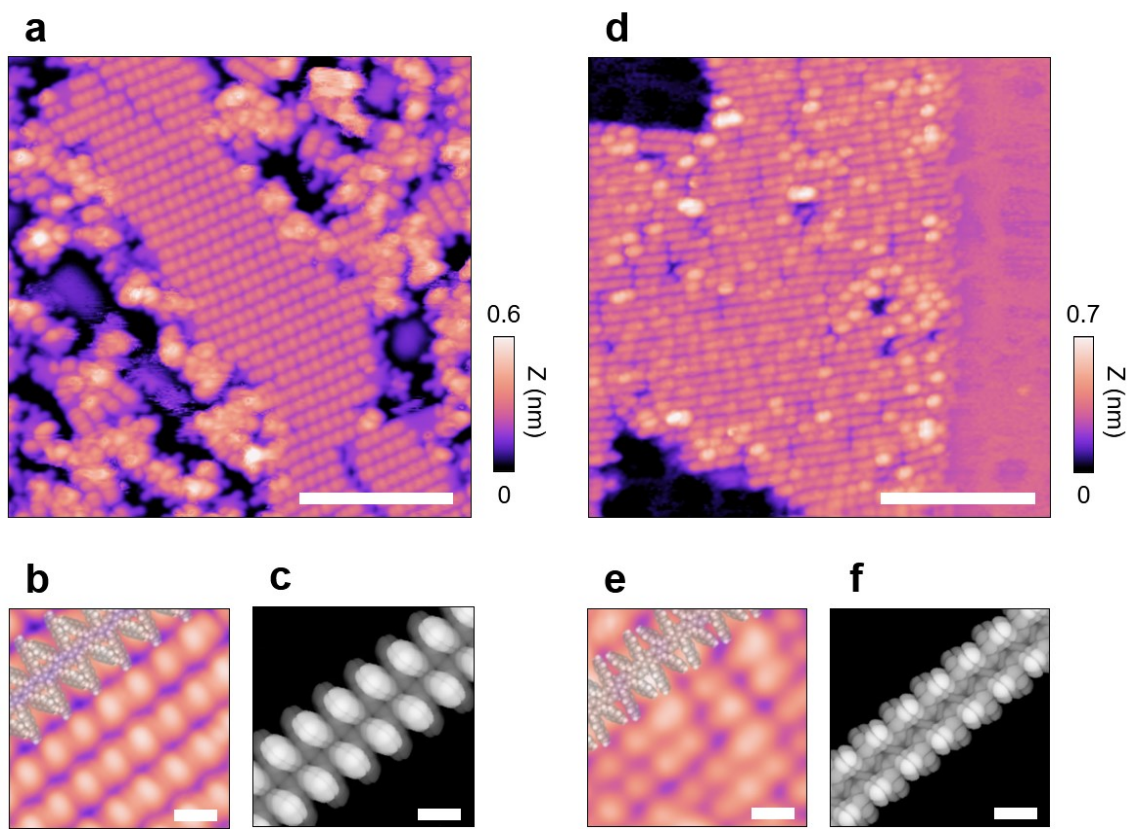

**Figure 2 | STM topographies and DFT simulations for 17- and 13-polymers.**
(**a**) Large-scale STM image (scanning paremeters: sample bias $V_s$ = 1.1 V, tunneling current $I_t$ = 10 pA) of 17-polymers on Au(111) and (**b**) small-scale image ($V_s$ = 1.1 V, $I_t$ = 10 pA). Structural model of single 17-polymer is superimposed in (**b**). (**c**) DFT simulated image (energy window $U$ = 1.1 eV) of the 17-polymer. (**d**) Large-scale STM image ($V_s$ = −1.2 V, $I_t$ = 70 pA) of 13-polymers on Au(111) and (**e**) small-scale image ($V_s$ = 1.0 V, $I_t$ = 10 pA) together with structural model. (**f**) DFT simulated image ($U$ = 1.0 eV) of the 13-polymer. All STM images were taken at 77 K. Scale bar, 10 nm in (**a, d**), 1 nm in (**b, c, e, f**).

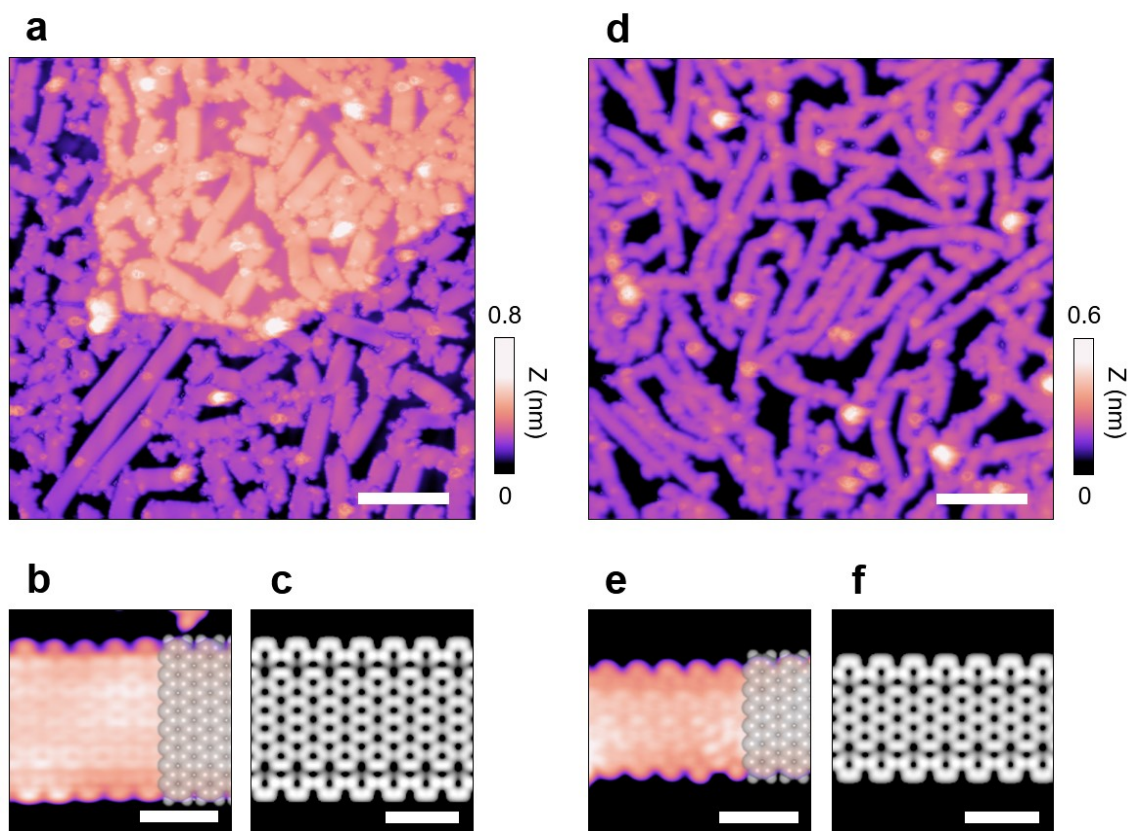

**Figure 3 | STM and DFT simulated images of 17- and 13-AGNRs.** (**a**) Overview STM image ($V_s = -1.6$ V, $I_t = 20$ pA) of 17-AGNRs on Au(111). (**b**) High-resolution STM image ($V_s = -1.4$ V, $I_t = 1.0$ nA) of single 17-AGNR together with structural model. (**c**) DFT simulated image ($U = -1.4$ eV) of the 17-AGNR. (**d**) Overview STM image ($V_s = -1.5$ V, $I_t = 20$ pA) of 13-AGNRs on Au(111). (**e**) High-resolution STM image ($V_s = -1.6$ V, $I_t = 1.2$ nA) of single 13-AGNR together with structural model and (**f**) corresponding DFT simulated image ($U = -1.6$ eV). All STM images were taken at 5 K. Scale bar, 10 nm in (**a, d**), 1 nm in (**b, c, e, f**).

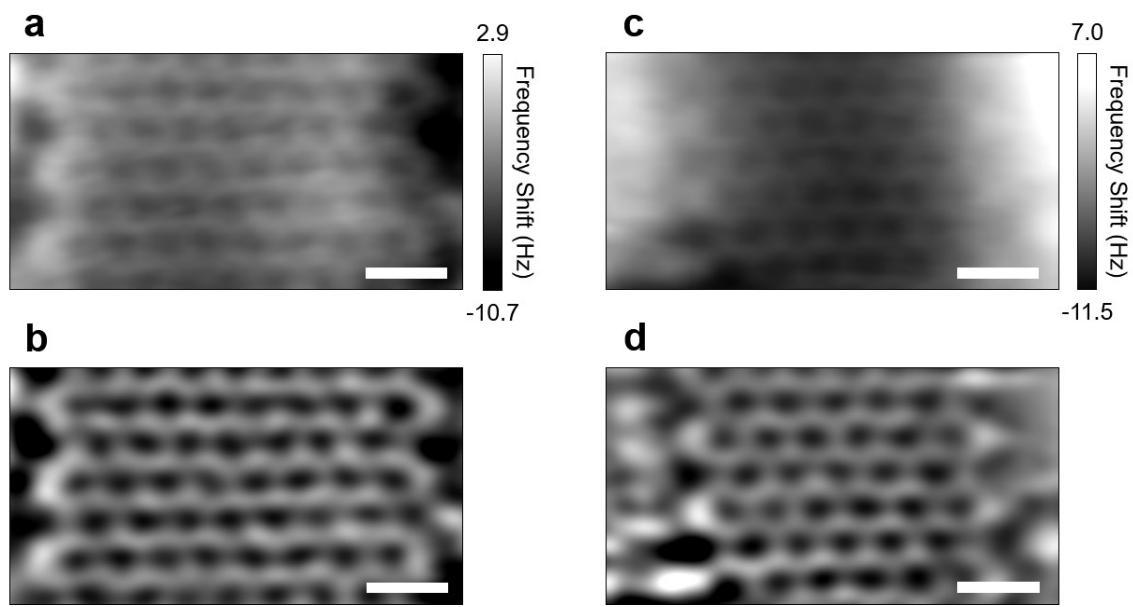

**Figure 4 | nc-AFM characterization of 17- and 13-AGNRs. (a)** Constant-height frequency shift image of single 17-AGNR measured by nc-AFM with a CO-functionalized tip. **(b)** Laplace filtered image of **(a)**. **(c)** Analogous frequency shift image of single 13-AGNR and **(d)** Laplace filtered image of **(c)**. Frequency shift images were taken at $V_s = 0$ V, oscillation amplitude of 0.10 nm and 5 K. All scale bars, 0.5 nm.

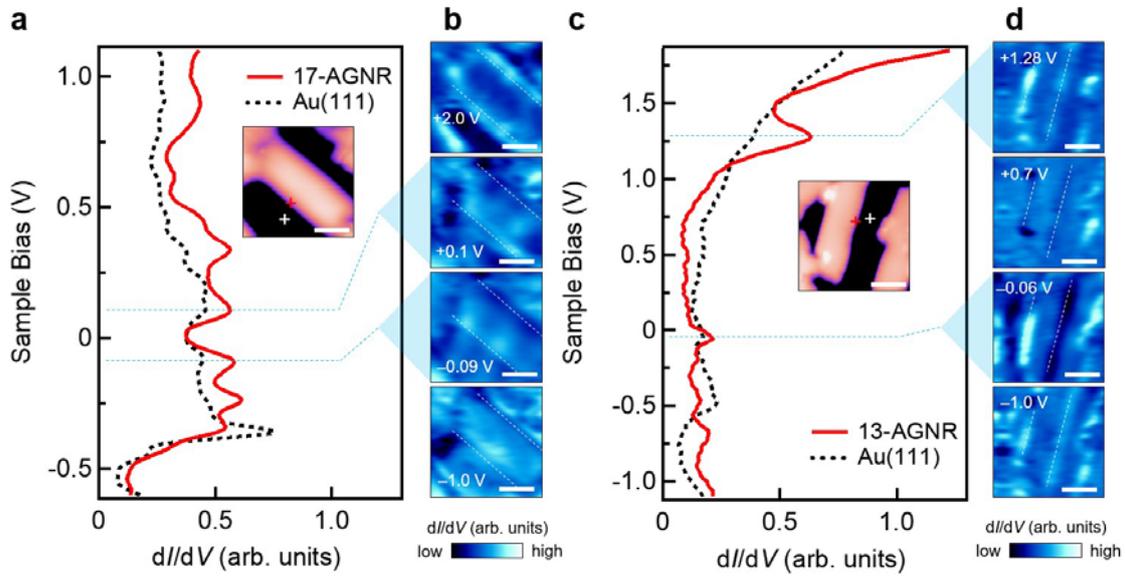

**Figure 5 | Electronic state characterization of 17- and 13-AGNRs.** (**a**) Differential conductance (d$I$/d$V$) point spectra recorded on edge of single 17-AGNR (red line) and Au(111) surface (black dotted line). Crosses in inset of STM image indicate tip positions for STS. Open-feedback setpoints were $V_s$ = −1.1 V, $I_t$ = 0.41 nA. (**b**) Constant-height d$I$/d$V$ maps ($I_t$ = 0.41 nA) of the 17-AGNR obtained at energies indicated in each map. Dashed lines indicate outer edges of the AGNR. (**c**) d$I$/d$V$ spectra of single 13-AGNR on Au(111) recorded at tip positions marked with crosses in inset STM image (setpoints: $V_s$ = 2.2 V, $I_t$ = 0.8 nA). (**d**) d$I$/d$V$ maps ($I_t$ = 0.8 nA) of the 13-AGNR obtained at respective energies shown in each map. All STS data were obtained at 5 K. All scale bars, 2 nm.

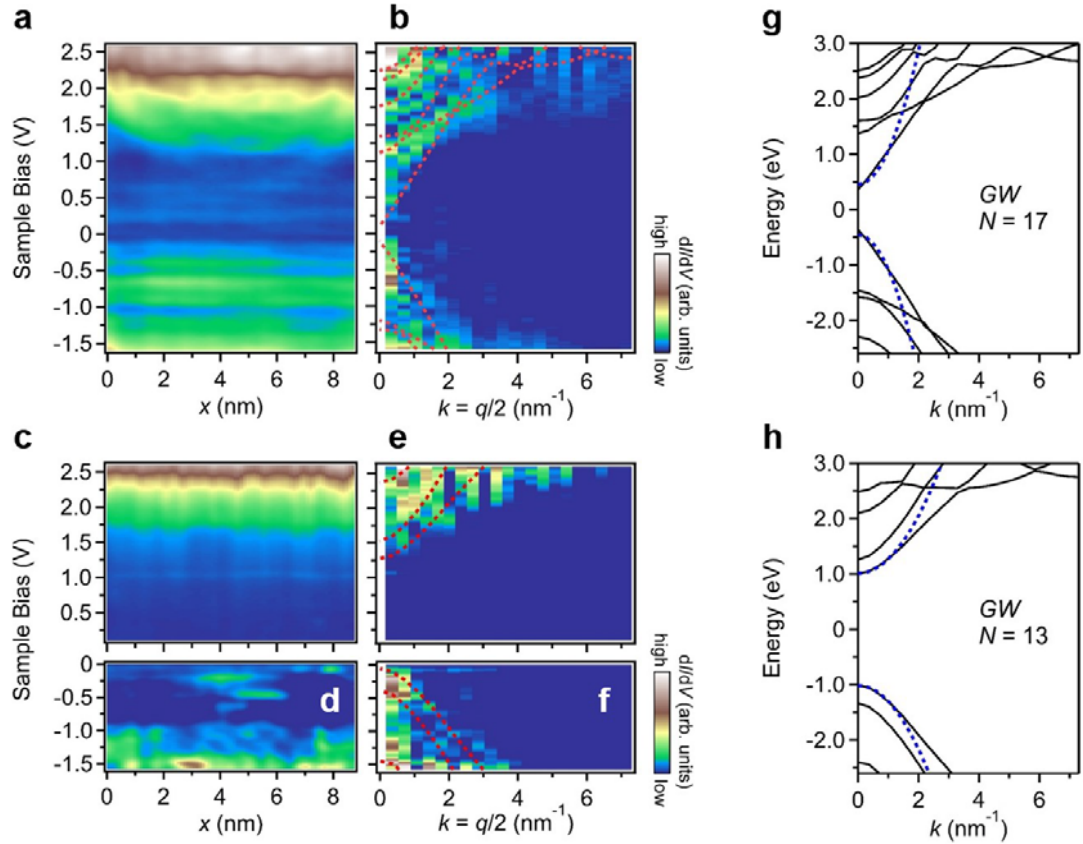

**Figure 6 | Electronic band structures of 17- and 13-AGNRs.** (**a**) Series of d$I$/d$V$ spectra obtained along one armchair edge of single 17-AGNR with length of 20$a$ on Au(111) (setpoints: $V_s$ = 2.6 V, $I_t$ = 0.7 nA, spacing $\delta x$ = 0.11). (**b**) Fourier-transformed map of (**a**) for $0 \leq k$ (= $q/2$) $\leq \pi/a$. (**c, d**) Analogous d$I$/d$V$ spectral maps of single 13-AGNR with length of 20$a$ on Au(111) for (**c**) unoccupied state (setpoints: $V_s$ = 2.6 V, $I_t$ = 0.7 nA, spacing $\delta x$ = 0.11) and (**d**) occupied state (setpoints: $V_s$ = −1.6 V, $I_t$ = 0.2 nA, spacing $\delta x$ = 0.11). (**e, f**) Fourier-transformed maps of (**c, d**). (**g, h**) *GW* quasiparticle band structures (black curves) of freestanding (**g**) 17- and (**h**) 13-AGNRs, aligned at center of gap (0 eV). Blue dotted curves show parabolic fits in rage of $0 \leq k \leq 1$ nm⁻¹ to obtain band effective masses. In Fourier-transformed maps of (**b**) and (**e, f**), red dotted curves indicate rigid-shifted quasiparticle bands of (**g**) and (**h**) to compare them with experimental band dispersions. All STS data were obtained at 5 K.

**Supplementary Information**

## Supplementary Note 1: Synthesis and characterization for BADBB and BNDBB

### 1. Materials and Methods

Reagents for synthesis were purchased from Wako, Nacalai Tesque and Sigma Aldrich, and were reagent-grade quality, obtained commercially, and used without further purification. Unless stated otherwise, column chromatography was carried out on silica gel 60N (Kanto Chemical, 40-50 μm). Analytical thin layer chromatography (TLC) was carried out on Art. 5554 (Merck, KGaA). Melting points (m.p.) were measured with a YAMAKO MP-J3. IR spectra were recorded on a JASCO FP-6600 and reported as wavenumbers $\nu$ in cm$^{-1}$ with band intensities indicated as s (strong), m (medium) and w (weak). The $^1$H NMR (600 MHz) and $^{13}$C NMR (150 MHz) spectra were recorded on a JEOL JNM-ECX 600 spectrometer and reported as chemical shifts ($\delta$) in ppm relative to TMS ($\delta = 0$). Broad peaks are marked as br. High-resolution MS was carried out on a MALDI-TOF (Bruker Autoflex II) or JEOL AccuTOF JMS-T100LC. X-ray crystallographic data were recorded at 90 K on a Bruker APEX II X-ray diffractometer equipped with a large area CCD detector by using graphite monochromated Mo Kα radiation ($\lambda = 0.71073$ Å).

### 2. Synthesis

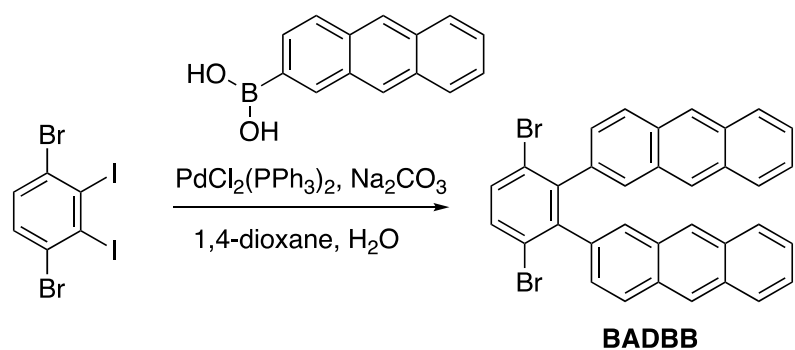

**BADBB**

1,4-dibromo-2,3-diiodobenzene[1] (300 mg, 0.62 mmol), 2-anthraceneboronic acid (282 mg, 1.27 mmol) and PdCl$_2$(PPh$_3$)$_2$ (45 mg, 0.064 mmol) were dissolved in 1,4-dioxane (15 ml), then 2 M Na$_2$CO$_3$ aqueous solution (6 ml) was added to the mixture. After argon bubbling for 45 min, the reaction mixture was stirred at 100 °C for 2 days. After cooling to room temperature, the solution was diluted with CH$_2$Cl$_2$, and washed with H$_2$O and brine, then the organic phase was dried over Na$_2$SO$_4$. The residue was concentrated *in* *vacuo*. The residue was purified by silica column chromatography ($CH_2Cl_2$/hexane = 1/10, $R_f$ = 0.05) to afford **BADBB** (242 mg, 67%) as a white solid.: m.p.: 245 °C; IR (KBr): 3089 (s), 3048 (m), 2958 (w), 1625 (w), 1533 (w), 1421 (w), 1301 (w), 1269 (w), 1154 (w), 1142 (w), 1018 (w), 952 (w), 909 (w), 890 (m), 869 (w), 814 (w), 800 (w), 738 (s), 665 (m); [1]H NMR (600 MHz, $CDCl_3$): 8.26 (s, 1 H), 8.20 (s, 1H), 8.19 (s, 1H), 8.17 (s, 1H),7.91 (d, $J$ = 1.8, 1H), 7.87–7.83 (br. m, 3H), 7.75–7.68 (m, 4H), 7.65 (s, 2H), 7.40–7.34 (m, 4H), 7.19–7.15 (m, 2H); [13]C-NMR (150 MHz, $CDCl_3$): 143.98, 143.85, 136.94, 136.77, 133.06, 132.97, 131.80, 131.70, 131.63, 130.73, 130.70, 130.28, 129.59, 128.89, 128.04, 128.02, 127.99, 127.97, 127.50, 127.45, 127.29, 127.20, 126.44, 126.41, 125.99, 125.97, 125.38, 125.36, 125.32, 123.68, 123.58; HRMS (Spiral) *m/z* = 585.9925, calcd for $C_{34}H_{20}Br_2$ = 585.9932.

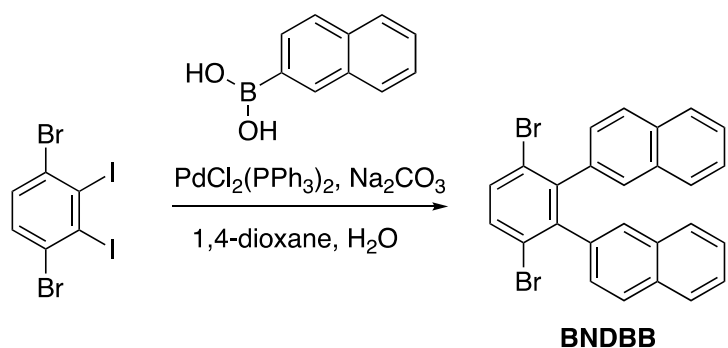

**BNDBB**

1,4-dibromo-2,3-diiodobenzene[1] (300 mg, 0.62 mmol), 2-naphthaleneboronic acid (206 mg, 1.2 mmol) and $PdCl_2(PPh_3)_2$ (45 mg, 0.02 mmol) were dissolved in 1,4-dioxane (15 ml), then 2 M $Na_2CO_3$ aqueous solution (6 ml) was added to the mixture. After argon bubbling for 45 min, the reaction mixture was stirred at 100 °C for 2 days. After cooling to room temperature, the solution was diluted with $CH_2Cl_2$ and washed with $H_2O$ and brine, then the organic phase was dried over $Na_2SO_4$. The residue was concentrated *in vacuo*. The residue was purified by silica column chromatography ($CH_2Cl_2$/hexane = 1/9, $R_f$ = 0.2 with hexane) to afford **BNDBB** (90 mg, 69%) as a white solid.: m.p.: 222 °C, IR (KBr): 3058 (m), 3042 (m), 2965 (w), 1599 (s), 1504(s), 1473 (s), 1425 (s), 1379 (w), 1364 (m), 1343 (w), 1264 (w), 1144 (s), 1126 (s), 1027 (s), 1017 (s), 961 (w), 892 (m), 856 (s), 826 (s), 816 (s), 803 (s), 769 (s), 748 (s); [1]H NMR (600 MHz, $CDCl_3$): 7.70–7.59 (m, 7 H), 7.55–7.53 (m, 3 H), 7.39–7.32 (m, 4 H), 7.16–7.13 (m, 2H); [13]C NMR (150 MHz, $CDCl_3$) 143.87, 143.82, 137.48, 137.40, 132.91, 132.88, 132.59, 132.53, 132.13, 129.19, 128.83, 127.98, 127.60, 127.55, 127.50, 127.23, 127.16, 125.99, 125.90, 127.71, 123.66; HRMS (Spiral) *m/z* = 485.9619, calcd for $C_{26}H_{16}Br_2$ = 485.9611.

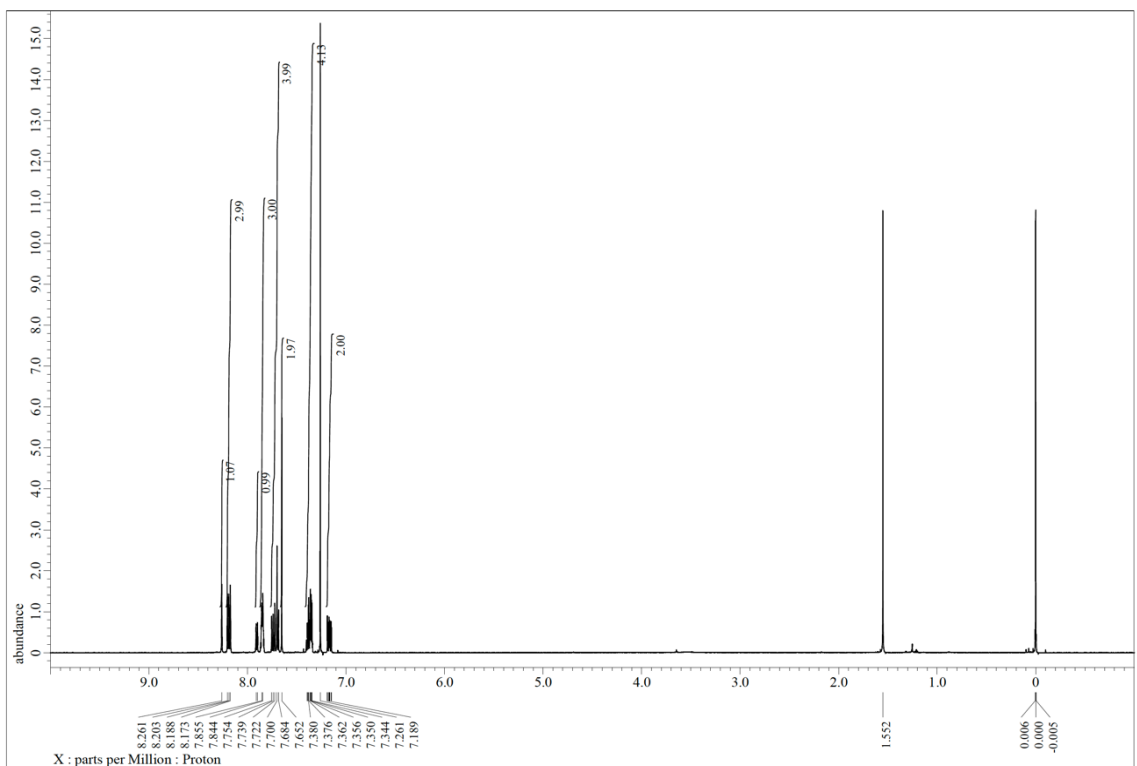

**Supplementary Figure 1 |** [1]H NMR spectrum of BADBB.

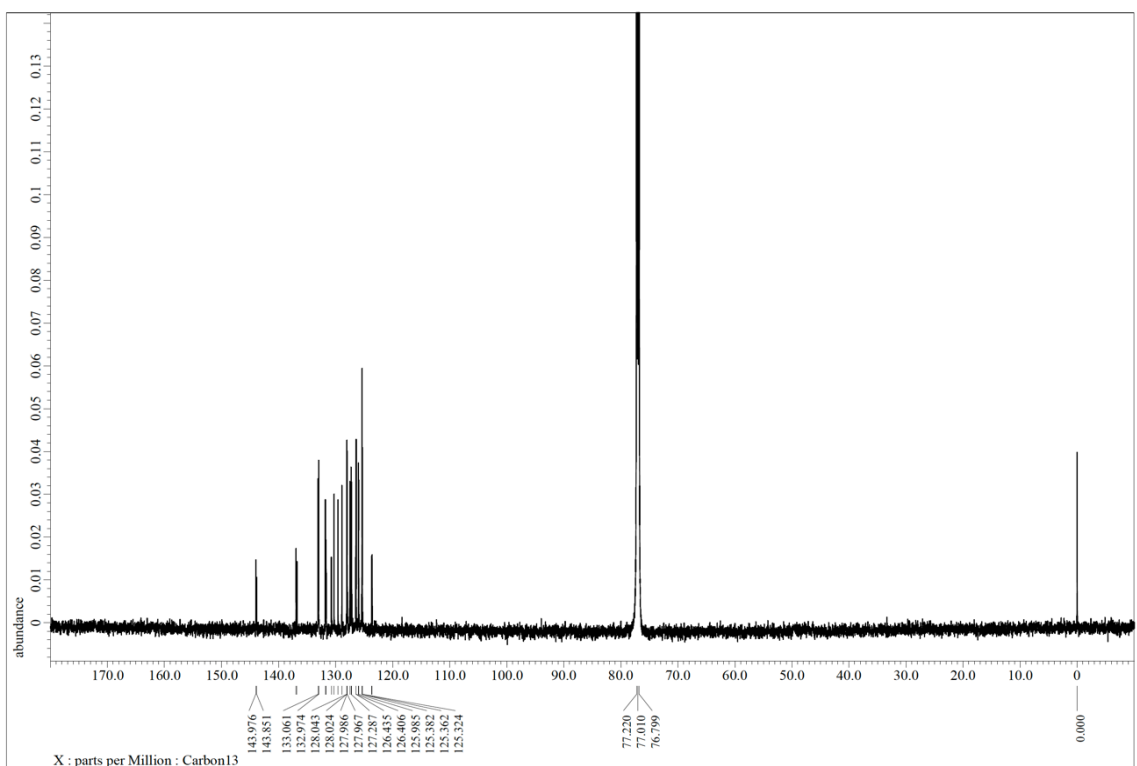

**Supplementary Figure 2 |** [13]C NMR spectrum of BADBB.

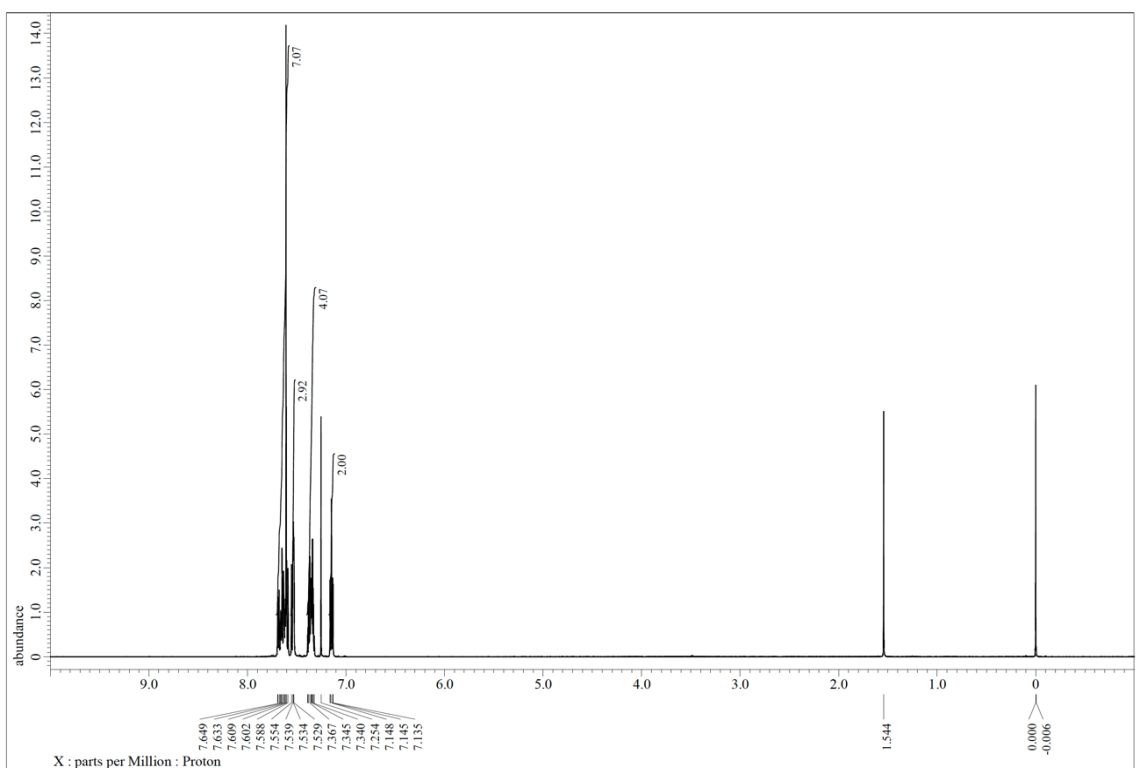

**Supplementary Figure 3 |** [1]H NMR spectrum of BNDBB.

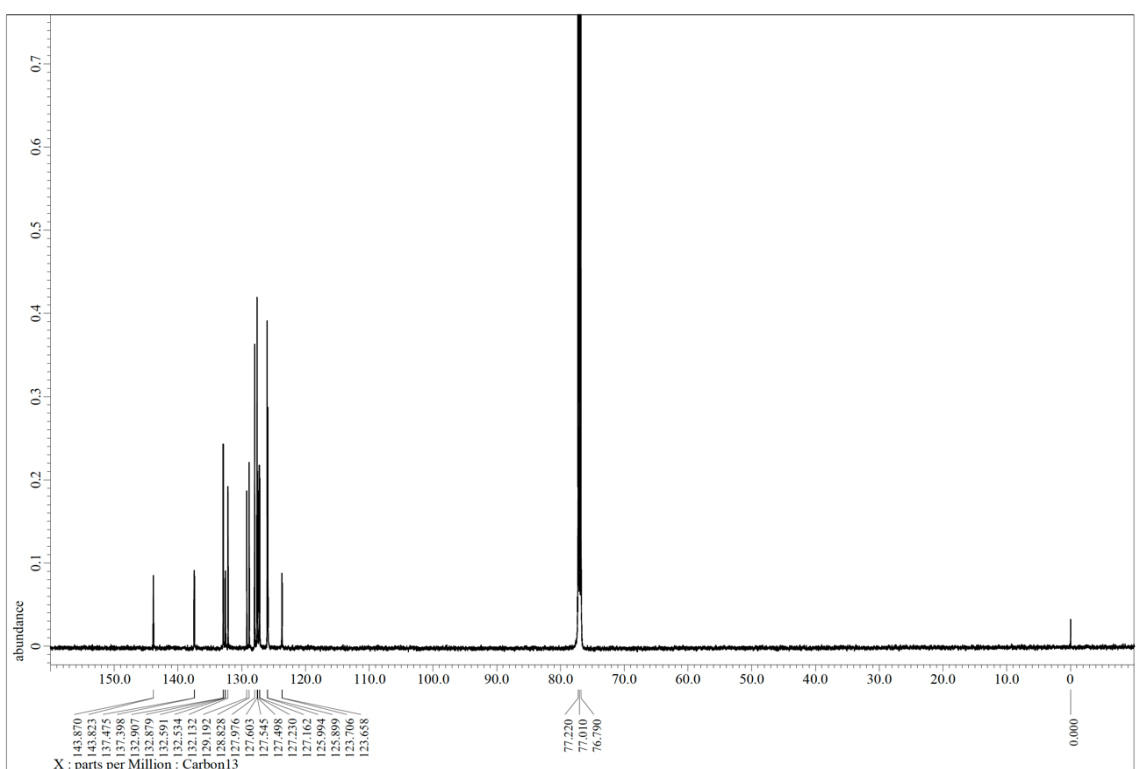

**Supplementary Figure 4 |** [13]C NMR spectrum of BNDBB.

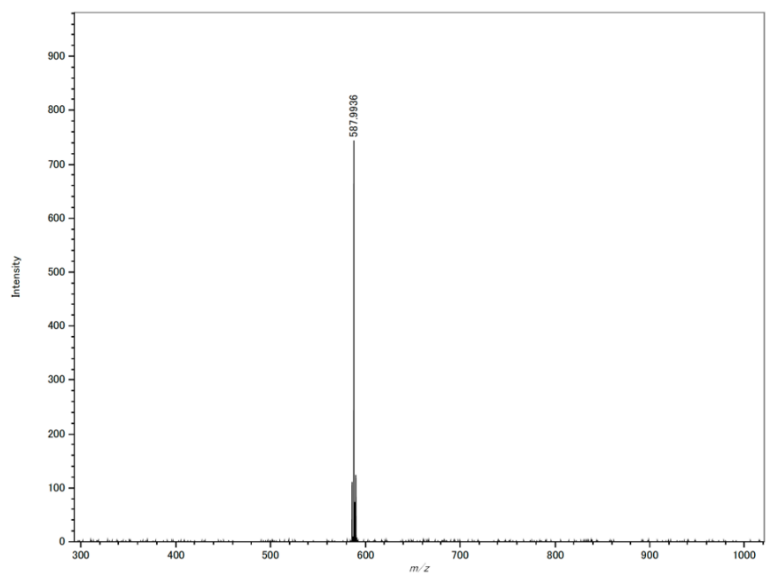

**Supplementary Figure 5 |** MS spectrum of BADBB.

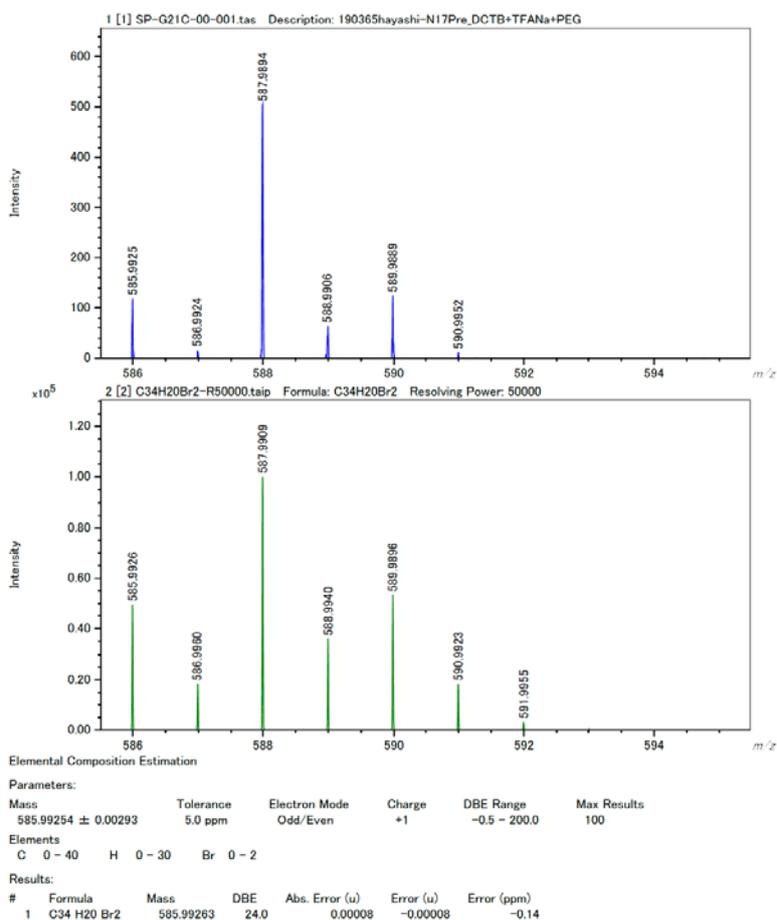

**Supplementary Figure 6 |** HRMS spectrum of BADBB.

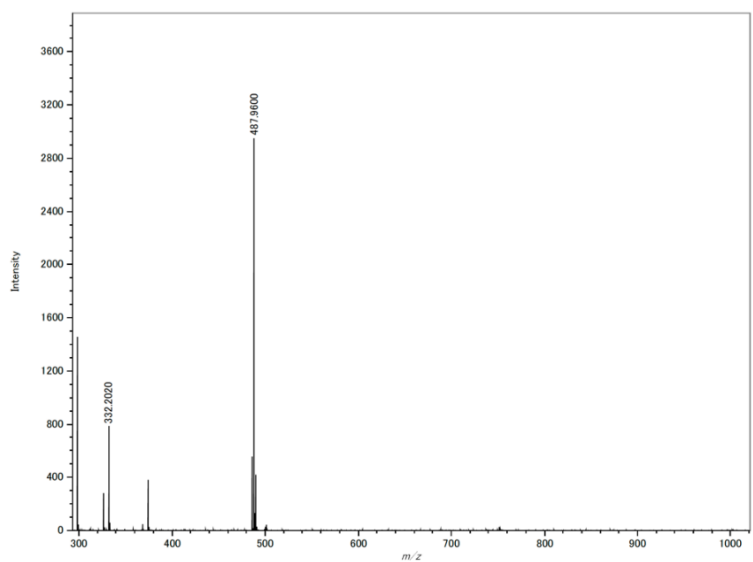

**Supplementary Figure 7 |** MS spectrum of BNDBB.

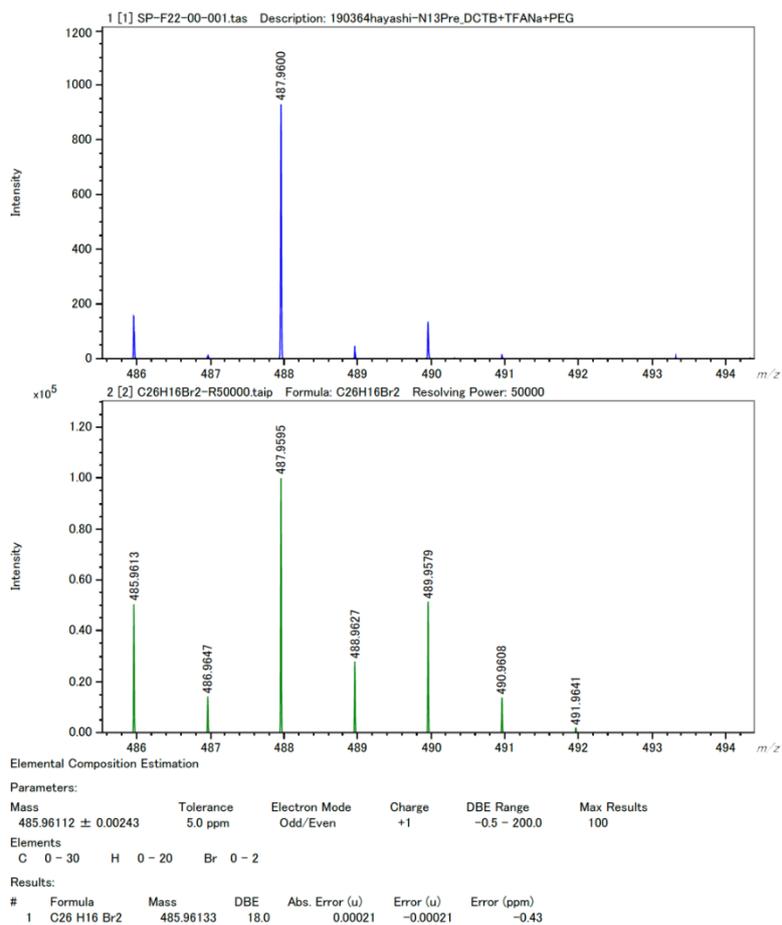

**Supplementary Figure 8 |** HRMS spectrum of BNDBB.

## 3. X-ray single-crystal analysis

**BADBB**

| | |
|---|---|
| Empirical formula | $C_{35}H_{22}Br_2Cl_2$ |
| Formula weight | 673.24 |
| Temperature | 90 K |
| Wavelength | 0.71073 Å |
| Crystal system | Orthorhombic |
| Space group | *Pnma* |
| Unit cell dimensions | $a$ = 23.217(13) Å $\quad\alpha$ = 90° |
| | $b$ = 19.569(11) Å $\quad\beta$ = 90° |
| | $c$ = 6.057(3) Å $\quad\gamma$ = 90° |
| Volume | 2752(3) Å$^3$ |
| $Z$ | 4 |
| Density (calculated) | 1.625 Mg/m$^3$ |
| Absorption coefficient | 3.165 mm$^{-1}$ |
| $F(000)$ | 1344 |
| Crystal size | $0.200 \times 0.010 \times 0.010$ mm$^3$ |
| Theta range for data collection | 1.754 to 24.498° |
| Index ranges | $-27 \leq h \leq 27$, $-22 \leq k \leq 22$, $-4 \leq l \leq 7$ |
| Reflections collected | 12842 |
| Independent reflections | 2356 [$R$(int) = 0.1644] |
| Completeness to theta = 24.498° | 100.0% |
| Absorption correction | Semi-empirical from equivalents |
| Max. and min. transmission | 0.969 and 0.389 |
| Refinement method | Full-matrix least-squares on $F^2$ |
| Data/restraints/parameters | 2356/0/179 |
| Goodness-of-fit on $F^2$ | 1.095 |
| Final $R$ indices [$I > 2\sigma(I)$] | $R_1$ = 0.0707, $wR_2$ = 0.1637 |
| $R$ indices (all data) | $R_1$ = 0.1385, $wR_2$ = 0.2095 |
| Extinction coefficient | n/a |
| Largest diff. peak and hole | 1.481 and $-1.064$ e.Å$^{-3}$ |

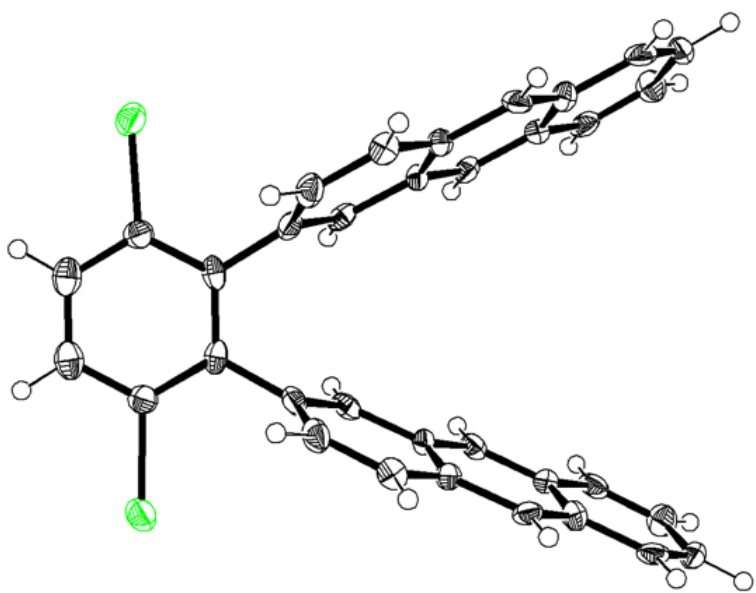

**Supplementary Figure 9 | Singe-crystal X-ray structure of BADBB.** Thermal ellipsoids represent 50% probability.

**BNDBB**

| | |
|---|---|
| Empirical formula | $C_{26}H_{16}Br_2$ |
| Formula weight | 488.21 |
| Temperature | 90 K |
| Wavelength | 0.71073 Å |
| Crystal system | Monoclinic |
| Space group | $C2/c$ |
| Unit cell dimensions | $a$ = 14.3226(14) Å |
| | $b$ = 20.4946(19) Å          $\beta$ = 114.665(2)° |
| | $c$ = 14.841(2) Å |
| Volume | 3959.0(8) Å$^3$ |
| $Z$ | 8 |
| Density (calculated) | 1.638 Mg/m$^3$ |
| Absorption coefficient | 4.104 mm$^{-1}$ |
| $F(000)$ | 1936 |
| Crystal size | $0.200 \times 0.050 \times 0.050$ mm$^3$ |
| Theta range for data collection | 1.853 to 25.999° |
| Index ranges | $-17 \leq h \leq 13, -23 \leq k \leq 25, -18 \leq l \leq 18$ |
| Reflections collected | 11393 |
| Independent reflections | 3895 [$R$(int) = 0.0372] |
| Completeness to theta = 24.498° | 99.8% |
| Absorption correction | Semi-empirical from equivalents |
| Max. and min. transmission | 0.821 and 0.736 |
| Refinement method | Full-matrix least-squares on $F^2$ |
| Data/restraints/parameters | 3895/0/253 |
| Goodness-of-fit on $F^2$ | 1.042 |
| Final $R$ indices [$I > 2\sigma(I)$] | $R_1$ = 0.0485, $wR_2$ = 0.0921 |
| $R$ indices (all data) | $R_1$ = 0.0628, $wR_2$ = 0.0973 |
| Extinction coefficient | n/a |
| Largest diff. peak and hole | 1.504 and $-1.110$ e.Å$^{-3}$ |

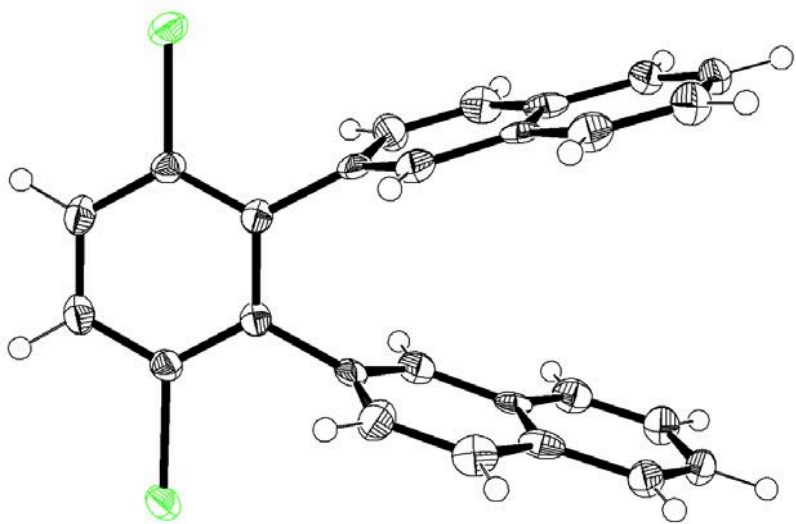

**Supplementary Figure 10 | Singe-crystal X-ray structure of BNDBB.** Thermal ellipsoids represent 50% probability.

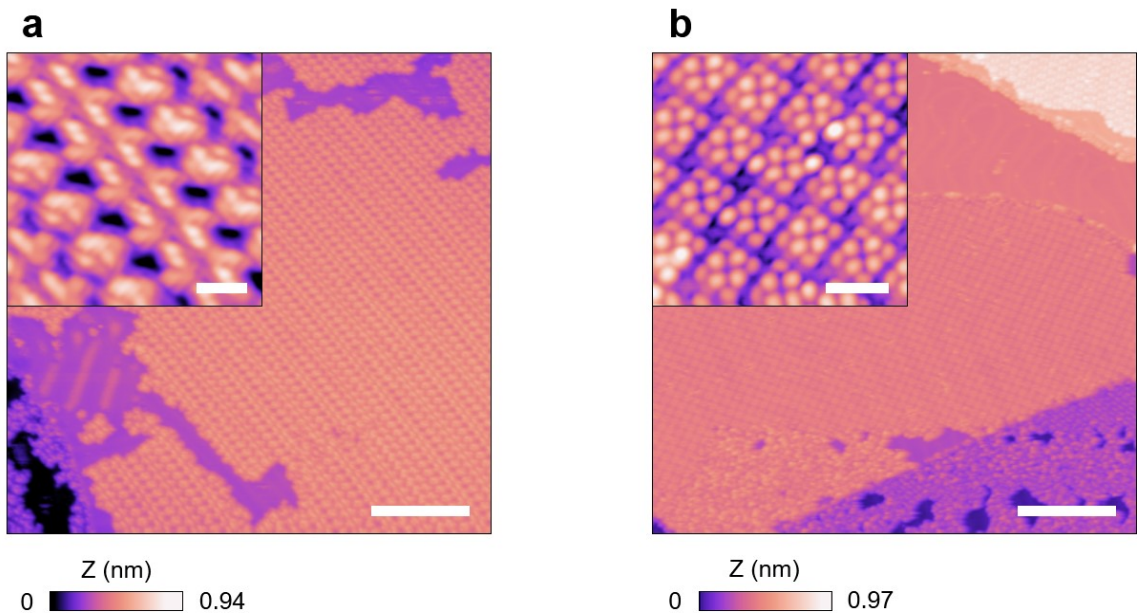

**a**

**b**

Z (nm)

0 ▬▬▬▬▬ 0.94

Z (nm)

0 ▬▬▬▬▬ 0.97

**Supplementary Figure 11 | STM topographic images of BADBB and BNDBB on Au(111) after room-temperature deposition.** (**a, b**) Large-scale STM images of (**a**) BADBB and (**b**) BNDBB. Scale bar, 20 nm. Insets, small-scale images in (**a**) and (**b**) observed within BADBB and BNDBB islands. Scale bar, 2nm. All images were taken at $V_s$ = 1.2 V, $I_t$ = 10 pA and sample temperature of 77 K.

## Supplementary Note 2: XPS measurements

The C 1$s$ core-level states of 17- and 13-AGNRs grown on Au(111)/mica substrates were investigated by X-ray photoelectron spectroscopy (XPS). XPS measurements were carried out with monochromated Al Kα radiation ($h\nu$ = 1,486.5 eV) at room temperature. The energy was calibrated with the Au Fermi edge, and the total energy resolution was 600 meV. Supplementary Figure 12a shows the C 1$s$ spectra of the 17- and 13-AGNRs. The main peaks are located at 284.2 eV in both AGNRs, and there are no tail structures derived from the oxygen-related components (C−O, C=O and COOH) in the higher binding energy side of the main peak. Note that while spectral shapes are almost the same in both types of AGNRs, the peak intensity of the 13-AGNRs is slightly higher than that of the 17-AGNRs in the lower biding energy side (indicated with an arrow). To quantitatively estimate the difference in the peak intensity, we carried out a numerical fitting of the C 1$s$ spectra. We assumed that the C 1$s$ peak consisted of C−C and C−H components. High-resolution photoemission studies using synchrotron radiation resolved two such components in the C 1$s$ peak of 7-AGNRs[2] and 9-AGNRs[3]. In the fitting procedure, we also assumed that the integrated intensity ratio between C−C and C−H components merely correspond to the ratio of the number of C atoms to C−C and C−H bonds in the unit cell (Supplementary Fig. 12b), namely, C−C : C−H = 30 : 4 for 17-AGNR and 22 : 4 for 13-AGNR. The fitting results are illustrated in Supplementary Fig. 12c, d. The determined peak positions are 284.2 and 283.6 eV for C−C and C−H components in both AGNRs, respectively. This numerical fitting can well reproduce the slight spectral difference between the 17- and 13-AGNRs.

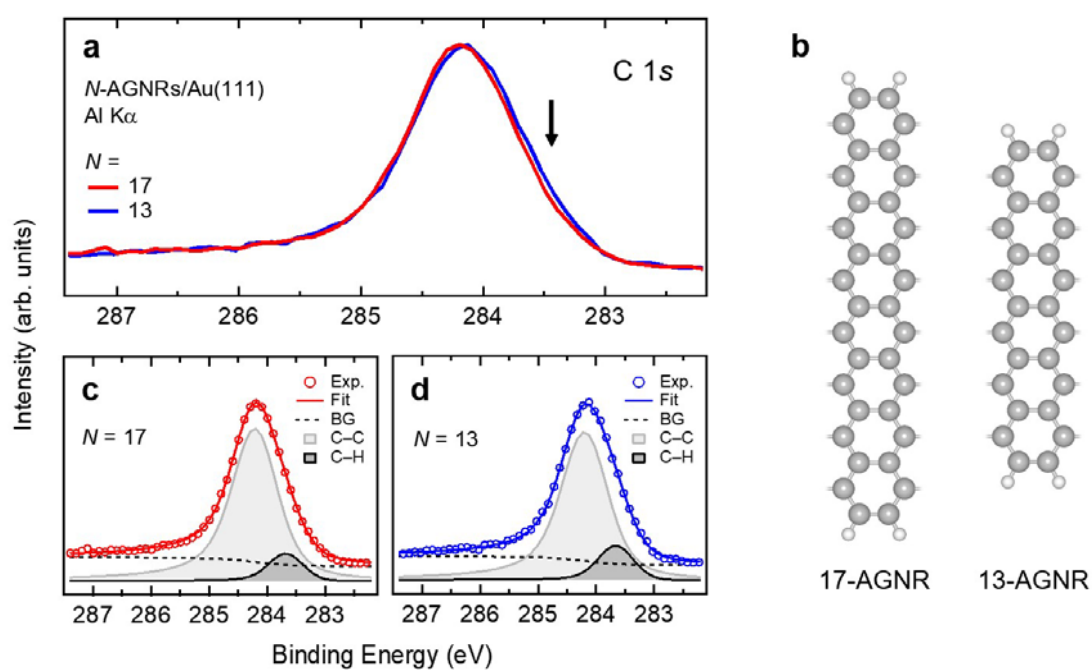

**Supplementary Figure 12 | XPS results of 17- and 13-AGNRs grown on Au(111)/mica.** (**a**) C 1$s$ core-level spectra. Spectra were normalized by peak height. (**b**) Unit cells of 17- and 13-AGNRs. (**c**, **d**) Numerical fitting to C 1$s$ spectra of (**c**) 17- and (**d**) 13-AGNRs.

## Supplementary Note 3: Raman spectroscopic measurements

We carried out Raman spectroscopy with a laser excitation wavelength of $\lambda_{\mathrm{ex}} = 633$ nm for the 17- and 13-AGNRs on Au(111)/mica. Supplementary Figure 13a shows the Raman spectra of the as-grown AGNRs. As a reference, the spectrum of the bare Au(111)/mica substrate is also shown. Although the ribbon-related $G$- and $D$-peaks are located at ~1,590 and ~1,330 cm$^{-1}$ in both AGNRs, respectively, these peak intensities are relatively weak compared to the background signals from the Au substrates. To make the Raman signals much more visible, we transferred the AGNRs onto SiO$_2$/Si substrates via a conventional transfer procedure[4]. The Raman spectra of the 17- and 13-AGNRs transferred onto SiO$_2$/Si are presented in Supplementary Fig. 13b. The spectra of both types of AGNRs exhibit not only prominent $G$- and $D$-peaks centered at 1,605 and 1,320 cm$^{-1}$, respectively, but also their overtone modes ($2D$ at 2,625 cm$^{-1}$, $D+G$ at 2,910 cm$^{-1}$ and $2G$ at 3,200 cm$^{-1}$). According to Raman simulations based on the density functional theory[1, 5], the frequency of the width-specific radial-breathing-like mode (RBLM) is predicted to be ~170 and ~220 cm$^{-1}$ for 17- and 13-AGNRs, respectively. To detect the RBLM signals, we carried out the Raman measurements with several excitation wavelengths ($\lambda_{ex}$ = 488, 633 and 785 nm), and the spectra in the RBLM frequency region are shown in Supplementary Fig. 13c. Nevertheless, the peaks derived from the RBLM do not appear in either AGNRs due to the off-resonant excitations. The RBLM peaks of 17- and 13-AGNRs would be detectable under the on-resonance Raman conditions by using lasers whose energies correspond to their optical gaps[1] and by using low-frequency Raman spectroscopy[6].

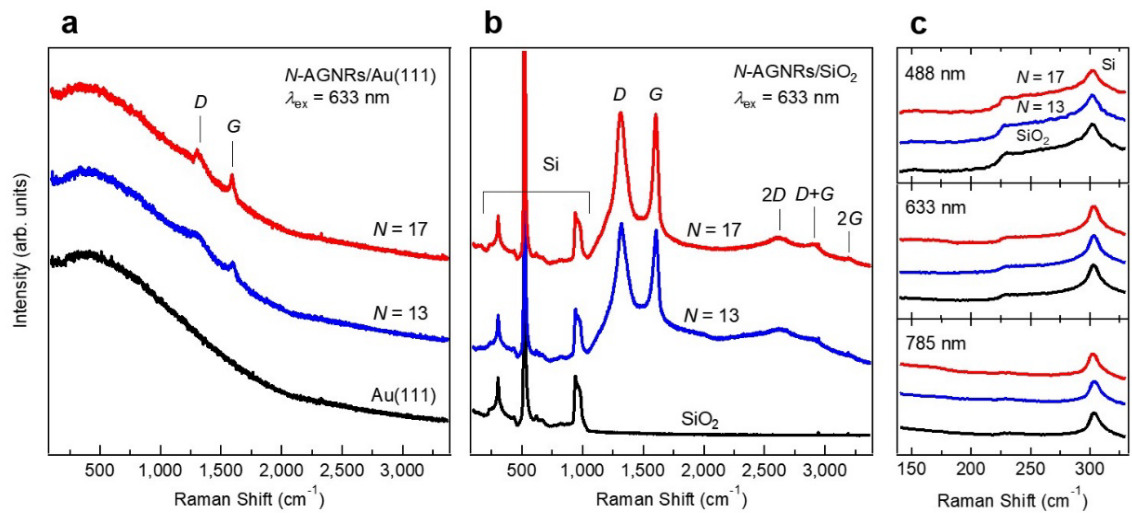

**Supplementary Figure 13 | Raman characterization of 17- and 13-AGNRs.** (**a**, **b**) Raman spectra obtained with $\lambda_{ex}$ = 633 nm of 17- and 13-AGNRs on (**a**) Au(111)/mica before transfer and (**b**) SiO$_2$/Si after transfer. (**c**) $\lambda_{ex}$-dependent Raman spectra of transferred samples in RBLM frequency region.

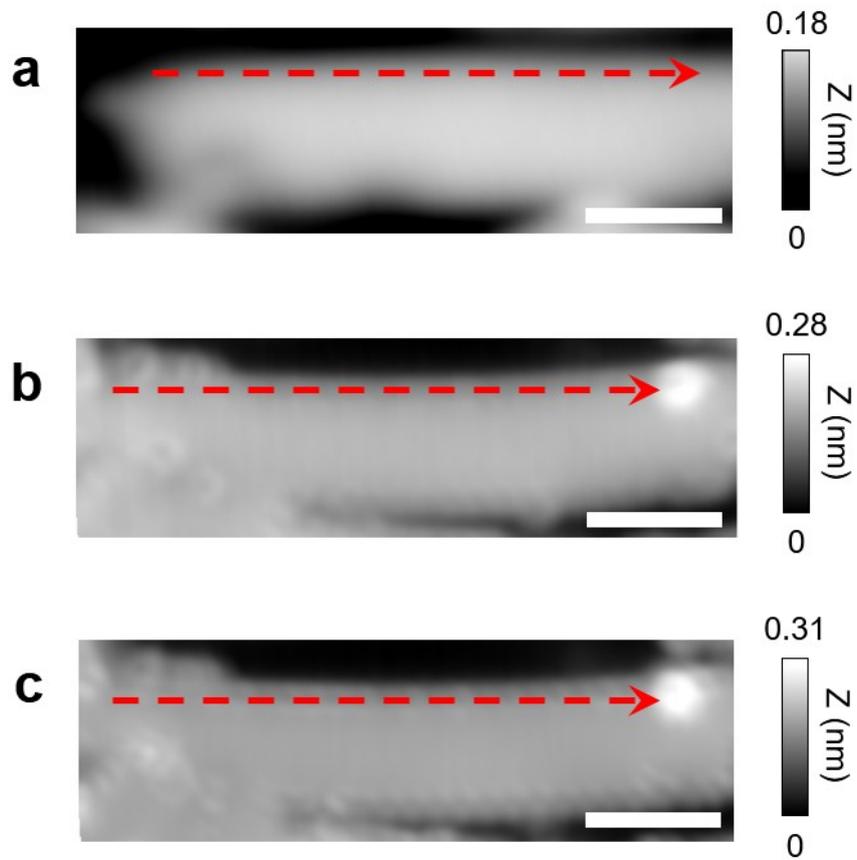

**Supplementary Figure 14 | 17- and 13-AGNRs used for FT-STS measurements.** (**a**) STM topographic image ($V_s$ = 2.6 V, $I_t$ = 0.7 nA) of single 17-AGNR. (**b**, **c**) STM images of single 13-AGNR taken at voltages in (**b**) unoccupied state ($V_s$ = 2.6 V, $I_t$ = 0.7 nA) and (**c**) occupied state ($V_s$ = −1.6 V, $I_t$ = 0.2 nA). d$I$/d$V$ spectral maps (Fig. 6a, c, d) were recorded along red dashed arrows in (**a**−**c**). All scale bars, 2 nm.